\documentclass[reprint,aps,pra,superscriptaddress,twocolumn,floatfix,lineno]{revtex4-1}
\usepackage{graphics,epsfig,amsfonts,amssymb,amsmath}
\usepackage{bbm}
\usepackage{color}
\usepackage{enumitem}
\usepackage{natbib}
\usepackage{hyperref}
\newcommand{\red}[1]{\textcolor{black}{#1}}

\newcommand{\ba}{\begin{array}{rl}}
	\newcommand{\ea}{\end{array}}
\newcommand{\dps}{\displaystyle}
\newcommand {\pd} [2] {\frac{\partial #1}{\partial #2}}

\begin{document}
% \linenumbers

%\author{}
%		%\email[]{Your e-mail address}
%		%\homepage[]{Your web page}
%		%\thanks{}
%		%\altaffiliation{}
%		\affiliation{}

\title{Theory of Hole-Spin Qubits in Strained Germanium Quantum Dots}
\author{L. A. Terrazos} 
\affiliation {Centro de Educa\c c\~ao e Sa\'ude, Universidade Federal de Campina Grande, Cuit\'e, PB 58175-000, Brazil}
\author{E. Marcellina} 
\affiliation{School of Physics, University of New South Wales, Sydney 2052, Australia}
\author{Zhanning Wang} 
\affiliation{School of Physics, University of New South Wales, Sydney 2052, Australia}
\author {S. N. Coppersmith}
\affiliation {Department of Physics, University of Wisconsin-Madison, Madison, Wisconsin 53706, USA}
\affiliation{School of Physics, University of New South Wales, Sydney 2052, Australia}
\author {Mark Friesen}
\affiliation {Department of Physics, University of Wisconsin-Madison, Madison, Wisconsin 53706, USA}
\author{A. R. Hamilton} 
\affiliation{School of Physics, University of New South Wales, Sydney 2052, Australia}
\author {Xuedong Hu}
\affiliation {Department of Physics, University at Buffalo, SUNY, Buffalo, New York 14260-1500, USA}
\author {Belita Koiller} 
\affiliation{Instituto de F\'{\i}sica, Universidade Federal do Rio de Janeiro, CP 68528, 21941-972 RJ, Brazil}
\author{A. L. Saraiva} 
\affiliation{Instituto de F\'{\i}sica, Universidade Federal do Rio de Janeiro, CP 68528, 21941-972 RJ, Brazil}
\author{Dimitrie Culcer} 
\affiliation{School of Physics, University of New South Wales, Sydney 2052, Australia}
\author {Rodrigo B. Capaz}
\affiliation{Instituto de F\'{\i}sica, Universidade Federal do Rio de Janeiro, CP 68528, 21941-972 RJ, Brazil}

\date{\today}

\begin{abstract}
We theoretically investigate the properties of holes in a Si$_{x}$Ge$_{1-x}$/Ge/ Si$_{x}$Ge$_{1-x}$ quantum well in a perpendicular magnetic field that make them advantageous as qubits, including a large ($>$100~meV) intrinsic splitting between the light and heavy hole bands, a very light ($\sim$0.05$\, m_0$) in-plane effective mass, consistent with higher mobilities and tunnel rates, and larger dot sizes that could ameliorate constraints on device fabrication. 
Compared to electrons in quantum dots, hole qubits do not suffer from the presence of nearby quantum levels (e.g., valley states) that can compete with spins as qubits.
The strong spin-orbit coupling in Ge quantum wells may be harnessed to implement electric-dipole spin resonance, leading to gate times of several nanoseconds for single-qubit rotations. 
The microscopic mechanism of this spin-orbit coupling is discussed, along with its implications for quantum gates based on electric-dipole spin resonance, stressing the importance of coupling terms that arise from the underlying cubic crystal field.
Our results provide a theoretical foundation for recent experimental advances in Ge hole-spin qubits.
\end{abstract}

\maketitle

\section{INTRODUCTION}

Hole spin qubits in strained germanium possess favorable properties for quantum computing, including
(1) the absence of valley degeneracy, which would otherwise compete with the spin degree of freedom for qubits formed in the conduction band of Si or Ge~\cite{friesen06},
(2) the high natural abundance of spin-0 nuclear isotopes in Ge, which may be further purified,
(3) the formation of hole states in $p$-type atomic orbitals whose wave function nodes occur at nuclear sites, suppressing unwanted hyperfine interactions~\cite{Bulaev:2007,de11}, and
(4) the very light in-plane effective mass~\cite{dobbie12,scappucci2018,Morrison:2016,Lodari2019}, allowing for larger dots and relaxing constraints on device fabrication.
The light mass also improves carrier mobilities, which can exceed $10^6$\,cm$^2$/V\,s for 2D Ge hole gases~\cite{dobbie12}.
Leveraging these strengths, rapid progress has been made in implementing high-fidelity one and two-qubit gate operations~\cite{Nadj-Perge:2010,yongjie12,Li2015,Liles2018,Vukui:2018,scappucci2018,Watzinger2018,Hardy:2019,HendrickxPreprint,Sammak2019,JirovecPreprint}.

Several of the most important advantages for qubits, such as the lifting of level degeneracy at the valence-band edge, the light effective mass, and access to Rashba spin-orbit coupling (SOC), which enables fast gate operations, are not available in the bulk. Rather, they emerge in SiGe/Ge/SiGe quantum wells due to confinement or strain. 

\red{While the main qualitative features of the electronic band structure of uniaxially strained germanium can be understood from simple $\bf k\cdot p$ theory, the approximation becomes less accurate with increasing strain and nanoscale confinement. A more quantitatively accurate approach requires treating the strain non-perturbatively, for example, by using \emph{ab initio} methods. Both approaches have advantages and are complementary. For example, $\bf k\cdot p$ theory allows us to exploit crystalline symmetries to simplify the calculations of the quantum dot wave functions, and it provides an accessible scheme for studying non-equilibrium dynamical evolution during qubit gate operations, such as operations based on electric-dipole spin resonance (EDSR). Moreover, in many cases, the results of \emph{ab initio} methods can be incorporated directly into $\bf k\!\cdot\! p$ theory to obtain more reliable results.}

In this work, we provide a theoretical foundation for the emergent physics of Ge quantum wells, and explanations for recent experimental observations, through detailed \emph{ab initio} band-structure calculations.
We gain further insight into the origins of qubit-friendly materials properties by performing $\bf k\!\cdot\! p$ calculations.
We place special emphasis on understanding the Rashba coupling, and the matrix elements connecting different orbital states.
Taken together, these ingredients enable electrically driven spin flips via \red{EDSR}, with fast, single-qubit gate \red{frequencies} of order 0.2~GHz.
In contrast with other recent work~\cite{HendrickxPreprint}, we propose here to exploit the large out-of-plane value of the Land\'{e} $g$-factor, so that relatively small external magnetic fields are needed for gate operation, making the qubit more compatible with superconducting gate structures, such as microwave resonators.
A large $g$-factor also helps to define the qubit with respect to thermal broadening.

\begin{figure}[t]
\centering
\includegraphics[width=2.7in]{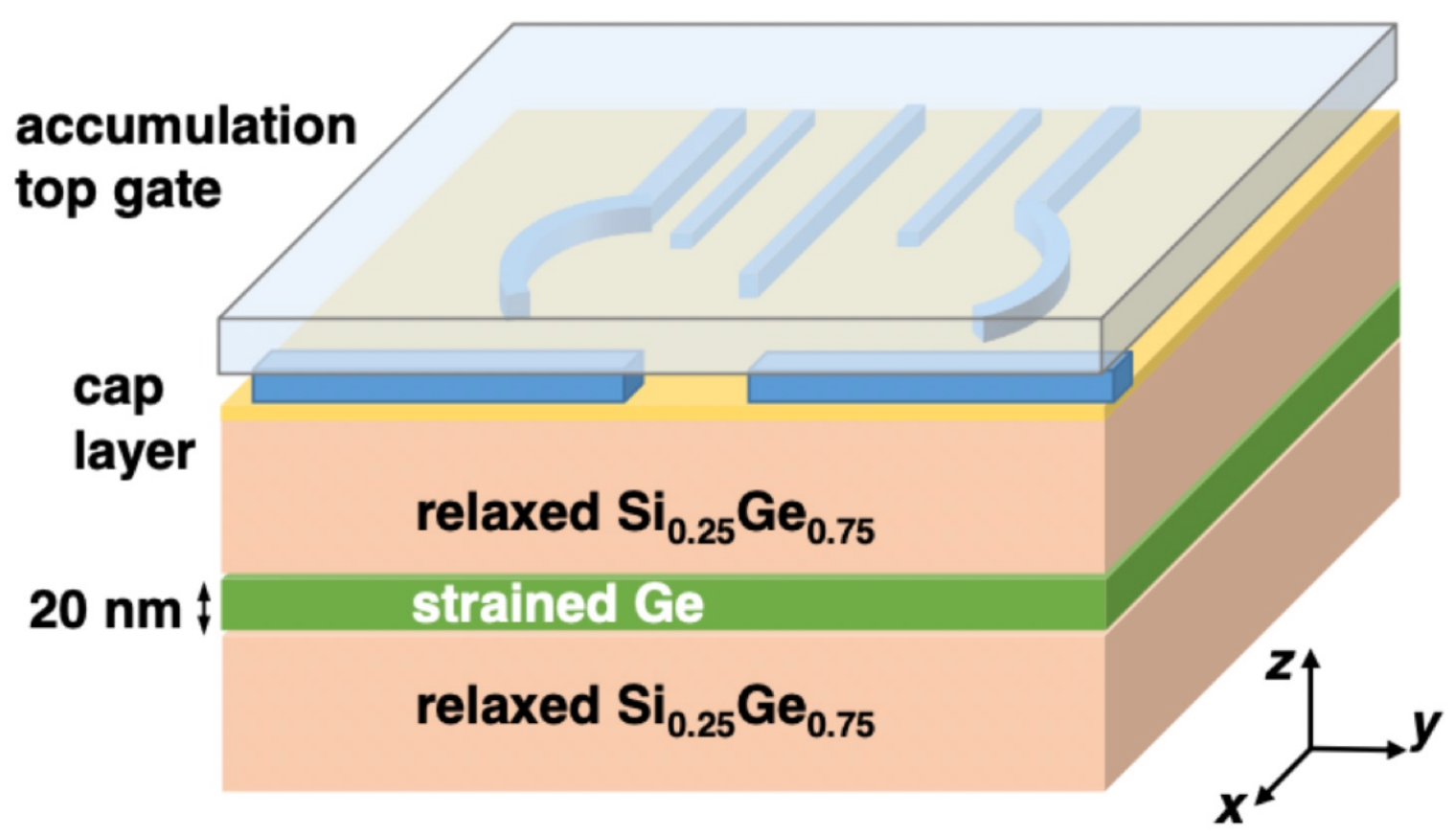}
\caption{
Cartoon depiction of a typical heterostructure and gate stack of a strained-Ge quantum well used to form hole-spin qubits in quantum dots.
Here, a 20~nm strained-Ge quantum well is grown epitaxially on a strain-relaxed Si$_{0.25}$Ge$_{0.75}$ alloy, as consistent with typical experiments~\cite{Sammak2019}.
For this arrangement, the strain in the Ge layer is $\varepsilon\approx-1$\%, as defined in Eq.~(\ref{eq:varepsilon_definition}).  
In addition to metal depletion gates (blue) and interspersed oxide layers (yellow), we assume a global top gate (transparent gray) that can accumulate a 2D hole gas in the quantum well in the absence of doping.
Here, $z$ is defined as the growth direction.}
\label{fig:device_schematic}
\end{figure}

\begin{figure}[t]
\includegraphics[width=\columnwidth]{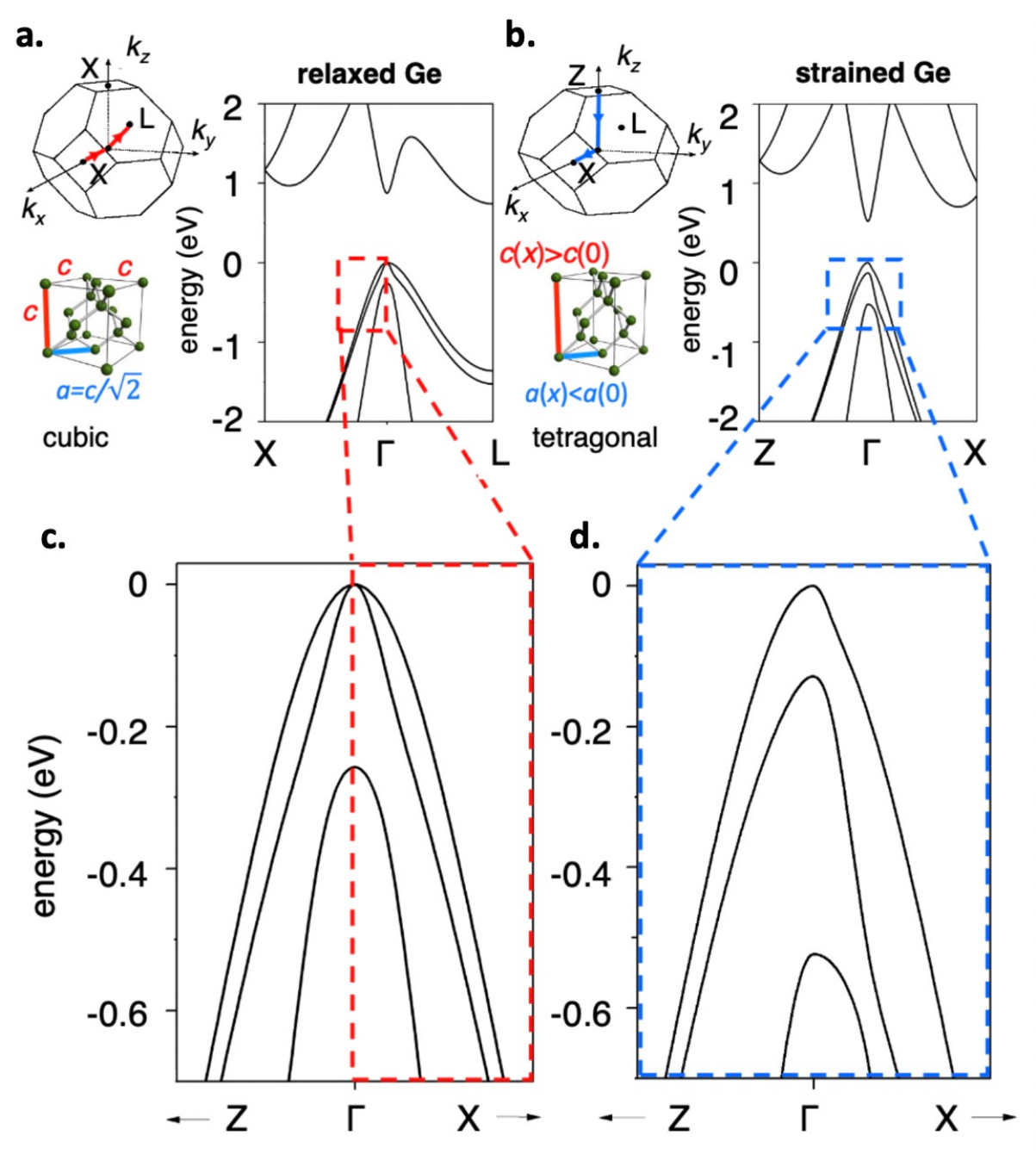}
\caption{
Electronic band structures for (a) relaxed vs.\ (b) uniaxially-strained Ge, obtained using DFT.
To the left of each plot we show the corresponding real and reciprocal space crystal structures (lower and upper diagrams, respectively), with lattice constants ($a$ and $c$) and symmetry points ($\Gamma$, X, Z, and L), as indicated.
[Note that the tetragonal deformation is exaggerated in (c), for clarity.] 
(c),(d) Blown-up band structures corresponding to (a) and (b).
Here, we focus on the $[100]$ ($x$) and $[001]$ ($z$) axes because of their relevance for quantum dot formation, and we note that [100] and [010] are equivalent.
In (a) and (c), cubic symmetry also makes the X and Z points equivalent and enforces a degeneracy between the top two hole bands at the $\Gamma$ point.
The lowest or ``split-off" band is completely detached from the others.
Away from the singular $\Gamma$ point, the hole bands are all doubly-degenerate.
In (b) and (d), the $x$-$z$ degeneracy is lifted and only the $x$-$y$ degeneracy remains.
The resulting band structure is highly anisotropic.}
\label{fig:old_fig1_bcde}
\end{figure}

The paper is organized as follows.
In Sec.~II, we describe the model system, including the heterostructure and top gates (Fig.~\ref{fig:device_schematic}).
In Sec.~\ref{sec:methods}, we provide technical details of the theoretical methods used in this work.
We summarize the \emph{ab initio} simulations of the quantum-well portion of the device and our $\bf k\!\cdot\! p$ Hamiltonian.
We describe our theoretical approach for modeling EDSR in two steps. 
We first outline a model for hole confinement in the vertical direction (perpendicular to the plane of the quantum well) and the lateral confinement of a quantum dot, and use this to obtain the effective Rashba spin-orbit Hamiltonian for our geometry. 
We then use this to determine the EDSR Rabi frequency when applying an in-plane ac electric field.
In Sec.~\ref{sec:results}, we describe the main results of our calculations, including the band-structure details obtained by \emph{ab initio} methods (Fig.~\ref{fig:old_fig1_bcde}), the corresponding in-plane and out-of-plane effective masses as a function of Ge concentration and strain (Fig.~\ref{dqdu3}), and the energy splittings between the valence bands (Fig.~\ref{fig:dqdu2_dft}).
We then apply $\bf k\!\cdot\! p$ methods to help clarify the origins of energy-level splitting, and the lifting of degeneracy, by artificially separating the effects of strain and SOC (Fig.~\ref{fig:dqdu2}).
Finally, we use our EDSR analysis to estimate the expected Rabi frequency for a realistic range of device parameters (Figs.~\ref{fig:color_maps} and \ref{fig:line_cut}).
In Sec.~\ref{sec:discussion}, we discuss our results and conclude by reviewing the predominant decoherence mechanisms for Ge hole qubits.

\red{\section{DEVICE STRUCTURE}}\label{sec:structure}
We consider a typical, electrically gated double-dot device such as the one schematically depicted in Fig.~\ref{fig:device_schematic}.
The essential features include a SiGe/Ge/SiGe heterostructure, an optional capping layer, and a set of patterned, nanometer-scale metal gates that are isolated from the heterostructure by oxide layer(s).
When sandwiched between strain-relaxed, Ge-rich SiGe alloys, the compressively strained Ge forms a type-I quantum well that can trap either electrons or holes~\cite{schaffler97}, although we focus exclusively on holes here.
Note that the details of the gate and oxide layers are unimportant for the following discussion.

For the heterostructure, we specifically consider an accumulation-mode gating scheme~\cite{angus07,eng15,zajac15} with no dopants. 
\red{The bottom SiGe barrier is grown with the same composition as the underlying SiGe virtual substrate, which is assumed to be strain-relaxed and dislocation-free. 
Next, a pure Ge quantum well is grown, epitaxially, atop the SiGe barriers, followed by another SiGe barrier layer, with the same composition as the bottom barrier.
The resulting quantum well is engineered to be compressively strained, with sharp Ge/SiGe interfaces on both sides~\cite{harame2004revolution,dobbie12,scappucci2018}.
The Si concentration in the SiGe barriers should be high enough to form a quantum well. 
For example, a strain-relaxed Si$_{0.25}$Ge$_{0.75}$ barrier yields a valence-band offset of $\sim$170~meV~\cite{schaffler97}, which is ample for trapping holes.}
The width of the well should be less than the critical thickness for forming additional dislocations; however, this is not typically a problem for Ge-rich alloys.
For example, the critical Ge thickness of a Si$_{0.25}$Ge$_{0.75}$ barrier is $\sim$30~nm~\cite{paul10}.
Finally, we note that Ge forms unstable oxides~\cite{su17} (similar problems also occur for SiGe alloys~\cite{legoues89}); it may therefore be beneficial to include a silicon capping layer, with a carefully chosen thickness~\cite{laroche16}.

\vspace{.1in}
\section{METHODS}\label{sec:methods}

\red{We investigate the electronic band structure of a strained Ge quantum well by considering two complementary theoretical approaches. 
We first compute the bulk properties of strained Ge using density functional theory (DFT). 
From these band structure calculations, we extract the relevant parameters for the $\bf k\!\cdot\! p$ approximation, which is also used to construct a Luttinger-Kohn-Bur-Pikus Hamiltonian (LKBP). 
The LKBP Hamiltonian incorporates the symmetries of the Bloch states and is used to characterize the spin-orbit structure of the hole bands.
We also use it to characterize EDSR, which enables rotations of the spin qubits.}

\subsection{Density Functional Theory}
\label{subsec:density_functional_theory}
 
Realistic, quantitative predictions for the band structure of strained Ge are key for assessing the viability of hole-spin qubits.
In this work, we compute the band structure using self-consistent, {\it ab initio} density functional theory (DFT), including spin-orbit interactions. \red{The calculations assume periodic boundary conditions, and therefore provide information about the bulk properties of strained Ge. We may then take into account effects associated with the quantum well and electrostatic confinement of the dot using simpler, semi-empirical approaches, such as effective mass theory.}

The calculations are performed using the full-potential linearized augmented plane wave method (FP-LAPW), as implemented in the Wien2k package~\cite{Blaha}.
Using the augmented plane wave plus local orbital (APW+lo) basis set~\cite{madsen2001,andersen1975,sjostedt2000}, the wave functions are expanded in spherical harmonics inside non-overlapping atomic spheres, with ``muffin-tin" radii $R_\mathrm{MT}$, and in plane waves for the rest of the unit cell (the interstitial region). 
In the present calculations we adopt $R_\mathrm{MT}$=0.95~{\AA } for Ge, and use 405 $\mathbf{k}$ points in the irreducible wedge of the Brillouin zone. 
For the spherical-harmonic expansion, the maximum orbital angular momentum is taken to be $l_\mathrm{max}$=10, while the plane-wave expansion in the interstitial region is extended to $k_\mathrm{max}$=9.0$/R_\mathrm{MT}$=9.47~\AA$^{-1}$, and the charge density is Fourier expanded up to $G_\mathrm{max}$=12~Ry. 
(These simulation parameters were all checked and found to yield numerical convergence.) 
Electron-electron interactions are described using the modified Becke-Johnson exchange potential + local density approximation (LDA) correlations~\cite{tran2009,becke2006}, which is known to yield accurate calculations of band gaps in semiconductors.

The primitive Bravais lattice used in our simulations is body-centered tetragonal with a two-atom basis consistent with the diamond structure. 
Details of the real and reciprocal lattice structure are depicted in the insets of Figs.~\ref{fig:old_fig1_bcde}(a) and \ref{fig:old_fig1_bcde}(b).
For unstrained Ge, the tetragonal lattice parameters are given by $a$=$b$=4.0008{\AA } in the plane of the quantum well, and $c$=${\sqrt{2}}a$=5.6580{\AA} in the growth direction. 
For a Si$_x$Ge$_{1-x}$ alloy with concentration $x$, the lattice constant $a(x)$ is modified, and if the quantum well is grown pseudomorphically, the same lattice constant is also imposed upon the strained Ge.
We define the compressive strain as 
\begin{equation}
    \varepsilon(x) =[a(x) - a(0)]/a(0)<0~.
    \label{eq:varepsilon_definition}
\end{equation}
For the SiGe alloy, Vegard's law then gives $\varepsilon(x) = -0.04 x$, while Poisson's ratio for germanium gives $\nu=0.27=-[c(x)-c(0)]/[a(x)-a(0)]$~\cite{wortman1965young}.
Combining these formulas yields an analytical expression for $c(x)$. 

The main results of our DFT calculations are reported in Sec.~\ref{subsec:results_DFT}.
To simplify the calculations, we do not explicitly consider a quantum-well geometry. 
Instead, we adopt a range of strain parameters consistent with a strained Ge quantum well sandwiched between strain-relaxed Si$_x$Ge$_{1-x}$ for the range $x\in [0,0.25]$.
From the previous discussion, this corresponds to compressive strains in the range $\varepsilon\in [-1,0]$ percent.

\subsection{$\mathbf{k\!\cdot\! p}$ Theory for Strained Germanium}

\red{Since quantum dots are large in comparison to the lattice parameter and are typically operated at low densities (ideally at the single-hole level), their localized wave functions can be expressed as superpositions of Bloch states centered near the $\Gamma$ point, $\mathbf{k}=0$. 
In this regime, it is common, and beneficial, to complement the DFT analysis with $\bf k\!\cdot\! p$ theory, a semi-empirical approximation that describes the band structure near the high-symmetry $\Gamma$ point. 
This approach provides physical intuition about the symmetries of the hole states and allows us to perform analytic calculations of the hole wave functions and dynamics. 
Of particular interest for our work, it gives insights into energy-splitting mechanisms associated with SOC and strain for the upper valence bands.
On the other hand, $\bf k\!\cdot\! p$ theory requires inputs from either first principles DFT calculations or experimental measurements. 
We now describe the details of our $\bf k\!\cdot\! p$ band-structure calculations.
The main results of these calculations are presented in Sec.~\ref{subsec:results_k_dot_p}.}

\red{A 6$\times$6 $\bf k\!\cdot\! p$ Hamiltonian describing the valence band states of a bulk, diamond-structure semiconductor was derived in Ref.~\cite{luttinger1955motion} by expanding a periodic electronic Hamiltonian in powers of the wave-vector components, $\mathbf{k}=(k_x,k_y,k_z)$, near the $\Gamma$ point. 
The allowable terms in this expansion are constrained by the symmetries of the crystal, which greatly simplifies the resulting Hamiltonian.}

\red{Similar symmetry arguments can also be used to determine the dependence of the Hamiltonian on the strain tensor elements, $\{\varepsilon_{ij}\}$. 
The framework we use for these calculations was developed by Bir and Pikus~\cite{bir1974}, who made use of the fact that the deformation potentials depend (approximately) linearly on the strain~\cite{van1989band}. 
We refer to the full model as the Luttinger-Kohn-Bir-Pikus (LKBP) Hamiltonian, which can be expressed in the notation of Ref.~\cite{chuang1995physics}, with the phase convention of Ref.~\cite{voon2009kp}, in} the basis of total angular momentum eigenstates,
 $|j,m_j\rangle \in \{|\frac{3}{2},\frac{3}{2}\rangle,|\frac{3}{2},\frac{1}{2}\rangle,|\frac{3}{2},-\frac{1}{2}\rangle,|\frac{3}{2},-\frac{3}{2}\rangle,|\frac{1}{2},\frac{1}{2}\rangle,|\frac{1}{2},-\frac{1}{2}\rangle\}$,  as
\begin{widetext}
\begin{equation}
\ba
H_\text{LKBP}\!=\!
\left(
\begin{array}{cccccc}
P+Q       &       -S       &       R       &       0     &  -S/\sqrt{2}   &  \sqrt{2} R  \\
-S^*      &       P-Q      &       0       &       R     &  -\sqrt{2} Q   &  \sqrt{3/2}S \\
R^*       &       0        &      P-Q      &       S     &  \sqrt{3/2}S^* &  \sqrt{2} Q  \\
0         &       R^*      &       S^*     &      P+Q    & -\sqrt{2} R^*  & -S^*/\sqrt{2}\\
-S^*/\sqrt{2}&-\sqrt{2} Q^*&  \sqrt{3/2}S  & -\sqrt{2} R &   P+\Delta     &    0         \\
\sqrt{2} R^* &\sqrt{3/2}S^*&  \sqrt{2}Q^*  & -S/\sqrt{2} &     0          &  P+\Delta  

\end{array}
\right),
\ea \label{eq:HLKBP}
\end{equation}
where 
\begin{equation}
\ba
&\dps P = P_k + P_{\varepsilon},\quad Q = Q_k + Q_{\varepsilon},\\[1ex]
&\dps R = R_k+R_{\varepsilon}, \quad S = S_k + S_{\varepsilon}. \label{eq:LKterms}
\ea
\end{equation}
Here, the $k$ subscripts refer to Luttinger-Kohn Hamiltonian matrix elements, \red{which reflect the bulk diamond structure and its symmetries}, defined as~\cite{Luttinger-1956-PR}
\begin{equation}
\ba
&\dps P_k=\frac{\hbar^2}{2 m_0}\gamma_1(k_x^2+k_y^2+k^2_z),\quad Q_k=-\frac{\hbar^2}{2 m_0}\gamma_2(2k^2_z-k_x^2-k_y^2),\\[1ex]
&\dps R_k=\sqrt{3}\frac{\hbar^2}{2 m_0}[-\gamma_2 (k_x^2-k_y^2)+2i\gamma_3 k_x k_y],\quad  S_k=\sqrt{3}\frac{\hbar^2}{m_0}\gamma_3 (k_x-i k_y)k_z.
\ea \label{eq:LKparameters}
\end{equation}
\red{When strain is introduced into the Luttinger-Kohn model, the unperturbed valence bands strongly hybridize.
The $\varepsilon$ subscripts in Eq.~(\ref{eq:LKterms}) refer to Bir-Pikus strain-matrix elements, defined as~\cite{bir1974}}
\begin{equation}
\ba
&\dps P_{\varepsilon}=-a_v (\varepsilon_{xx}+\varepsilon_{yy}+\varepsilon_{zz}),\quad 
\dps Q_{\varepsilon}=-\frac{b_v}{2}(\varepsilon_{xx}+\varepsilon_{yy}-2\varepsilon_{zz}),\\[1ex]
&\dps R_{\varepsilon}=\frac{\sqrt{3}}{2}b_v(\varepsilon_{xx}-\varepsilon_{yy})-id\varepsilon_{xy},
\quad  S_{\varepsilon}=-d_v(\varepsilon_{xz}-i\varepsilon_{yz}).
\ea \label{eq:Pikus}
\end{equation}
\end{widetext}
\red{These strain elements also reflect the underlying lattice symmetries, as seen in the form of the strain-tensor elements, $\varepsilon_{ij}$, which mirror the $k_i k_j$ terms appearing in the Luttinger-Kohn parameters, Eq.~(\ref{eq:LKparameters}).} 
The Pikus-Bir expressions in Eq.~(\ref{eq:Pikus}), are generic, and we note that the parameter $\varepsilon_{xx}$(=$\varepsilon_{yy}$) is equivalent to $\varepsilon(x)$, defined in Eq.~(\ref{eq:varepsilon_definition}).
In this work, we focus on the special case of uniaxial strain, for which $\varepsilon_{zz}$=$-2(C_{12}/C_{11})\varepsilon_{xx}$ and $\varepsilon_{xy}$=$\varepsilon_{yz}$=$\varepsilon_{zx}$=0, leading to $R_{\varepsilon}$=$S_{\varepsilon}$=0.

\red{We note that results similar to those reported here can be obtained from alternative starting points, such as Kane's model, which includes the lowest conduction band, in addition to the upper valence bands~\cite{voon2009kp}.
However, the LKBP model is commonly adopted when the conduction band is not of direct interest.
The resulting band curvatures, non-parabolicities, and energy splittings from the LKBP model closely mirror those obtained from Kane's model.}

\red{The key ingredients for describing physics of quantum dots are all contained in Eq.~(\ref{eq:HLKBP}).
For example, $\Delta$ is the energy splitting between the topmost valence bands and the split-off band at $k=0$, in the absence of strain.
The strain is captured by the parameters $\{\varepsilon_{ij}\}$ in the Bir-Pikus expressions, and the quantum confinement due to gate-induced electric fields, as well as the quantum-well band offsets, is captured by the wave vectors, $\{k_i\}$.}

The essential physical parameters in Eqs.~(\ref{eq:HLKBP})-(\ref{eq:Pikus}) include the bare electron mass, $m_0$, the Luttinger parameters~\cite{Luttinger-1956-PR}, $\gamma_1$=13.38, $\gamma_2$=4.24, and $\gamma_3$=5.69, the deformation potentials~\cite{fischetti1996band}, $a_v$=$2.0$~eV, $b_v$=$-2.16$~eV, and $d_v$=$-6.06$~eV, and the elastic stiffness constants for the strain-stress tensor~\cite{wortman1965young}, $C_{11}$=129.2~GPa and $C_{12}$=47.9~GP.
In this work, we adopt the experimentally measured energy splitting of the split-off band, for bulk, relaxed Ge~\cite{schaffler97}, $\Delta$=0.296~eV.

\subsection{Calculating the Rashba Spin-Orbit Coupling}\label{sec:Rashba}
\noindent

\red{We now consider the practical consequences of the strained band structure for qubit implementations, which will be used in the following section to estimate EDSR driving speeds.
We envision a single-hole spin qubit in an electrostatically defined quantum dot, formed in a Ge quantum well.
Due to the inversion asymmetry inherent in the approximately triangular well we consider, Rashba SOC is expected to be exceptionally strong. 
We note that the Rashba effect is purely two-dimensional (2D), and although it depends on the energy splitting of the split-off band, $\Delta$, its physical origins are distinct.
Since Ge has near-inversion symmetry, the Dresselhaus interaction is known to be absent in the bulk. 
In quantum wells, Dresselhaus-like terms may arise due to the presence of interfaces~\cite{Durnev_PRB2014}. 
However, these depend on the coupling to the conduction and split-off bands and we expect them to be quite weak in a quantum well, because the hole wave functions barely enter into the barrier regions ~\cite{Marcellina2017}. 
Moreover, when Rashba terms are present in hole systems they tend to overwhelm all other spin-orbit interactions~\cite{Marcellina2017}.
With these observations in mind, we focus only on the Rashba SOC for the remainder of this work, and discuss the ways it can be used to implement electric-dipole spin resonance for qubit rotations and spin dipole-dipole entanglement.}

There are two prerequisites for observing Rashba SOC in a quantum well: a broken structural symmetry and an intrinsic SOC.
The broken symmetry is provided here by an asymmetric confinement potential of the form
\begin{equation}
V_z(z) = \left\{
  \begin{array}{cl}
    e F_z z & (|z|<d/2) \\
    \infty & (\text{otherwise}) 
  \end{array}
\right. , \label{eq:triangle}
\end{equation}
where $F_{z}$ is the average electric field across the quantum well, and the well width, $d$=20~nm, is held fixed for all our calculations.
(Note that $d$ is not an important parameter in this calculation, since the electric field draws the hole wave function to the top of the quantum well, so that it does not interact strongly with the bottom of the well.)
The total Hamiltonian for the vertical confinement of holes is then given by
\begin{equation}\label{4H}
H_z=
H_\mathrm{LKPB}(k^2,\hat{k}_z)+V_z(z),
\end{equation}
where $\hat{k}_z= -i\pd{}{z}$ and $k^2=k_x^2+k_y^2$.

\red{The strength of the Rashba SOC depends on the details of the wave function confinement and on the local electrostatics.
It is mainly determined by the hybridization of the top two valence bands, since the split-off band is far away in energy.
However, the split-off band indirectly enters the calculation because it affects other parameters, such as the effective masses and the intrinsic splitting of the top bands. 
To estimate the confinement along the $\hat{\mathbf{ z}}$ direction, we therefore consider the full three-band model (not counting spin), as described by $H_\text{LKPB}$, which includes both strain and SOC effects.}
We introduce variational, effective-mass wave functions for each of the bands, given by~\cite{Bastard-1983-PRB}: 
\begin{equation}\label{Bastard}
\varphi_i(z) =
\left\{
  \begin{array}{cl}
  \frac{\sin\left[\frac{\pi}{d}\left(z + \frac{d}{2}\right)\right] \exp \left[ -b_i \left(\frac{z}{d} + \frac{1}{2} \right) \right]}{\pi \sqrt{d\frac{\exp(-b_i)  \sinh(b_i) }{2 \pi^2 b_i + 2 b_i^3}}} & (|z|<d/2) \\
  0 & (\text{otherwise})
  \end{array}
\right. .
\end{equation}
Here, $i$ is the band index and $\{b_i\}$ are the dimensionless variational parameters. \red{Physically, the ratios $d/b_i$ represent the effective widths of the wave functions.
A separate variational parameter is required for each of the bands because of their distinct effective masses.}
We determine their values by minimizing the eigenvalues of $H_z$ in the limit of $k=0$, in which case the Hamiltonian is already diagonal and the bands decouple.   
The effective Rashba coupling within the topmost band is determined by applying a Schrieffer-Wolff transformation to Eq.~(\ref{4H}), using the states shown in Eq.~(\ref{Bastard}), to eliminate the coupling to the other two bands~\cite{Marcellina2017}.
In this way, we obtain the effective Hamiltonian $H_0 + H_R$, where $H_0 = \hbar^2(k_x^2+k_y^2)/2m_x$ is the kinetic energy in the effective mass approximation.
\red{For electrons, the Rashba interaction couples states with $\Delta m_j=\pm 1$. In contrast, the topmost valence band is doubly degenerate, with $|m_j|=3/2$. Transitions within this band therefore require that $\Delta m_j=\pm 3$, consistent with Hamiltonian operators that are predominantly cubic in $k$~\cite{Marcellina2017,WinklerBook}, and may be expressed as}
\begin{multline} \label{eq:HSO} % Two lines
H_R = i\alpha_{R2}(k_+^3\sigma_- - k_-^3\sigma_+)  \\
+i\alpha_{R3} \, (k_+k_-k_+\sigma_+ - k_-k_+k_-\sigma_-),
\end{multline}
where $\sigma_\pm$=$\sigma_x \pm i\sigma_y$ are Pauli spin matrices and $k_\pm=k_x\pm ik_y$.
The coupling constants $\alpha_{R2}$ and $\alpha_{R3}$ are derived in Ref.~\cite{Marcellina2017}. 
Here, $\alpha_{R2}$ arises from the spherically symmetric component of the Luttinger-Kohn Hamiltonian, while $\alpha_{R3}$ arises from the cubic-symmetric component. 

\subsection{Calculating the EDSR Rabi Frequency}
\label{sec:Rabi_calculations}

\red{An external magnetic field is used to define the quantization axis of the spin qubit.
This field also generates rotations about the qubit's $\mathbf{\hat z}$ axis.
However, a universal gate set also requires being able to perform rotations about the $\mathbf{\hat x}$ axis, using a technique such as spin resonance.
To implement electric-dipole spin resonance, microwave voltage signals are brought to the qubit through the top-gate electrodes used to confine the hole laterally and form the quantum dot. This time-varying drive causes the hole to oscillate in the plane of the quantum well. SOC then provides a mechanism for converting the orbital motion into spin oscillations~\cite{tokura06,obata10}. When the drive frequency is resonant with the spin precession frequency, the desired $x$ rotations occur. We now estimate the resulting gate speed.}

\red{We assume the presence of Rashba SOC, as described in the previous section.
Contrary to other proposals that we have seen, we assume the quantizing $B$-field is oriented perpendicular to the plane of the quantum well, to take advantage of the large out-of-plane $g$-factor~\cite{WinklerBook}, $g_z$, which reduces the constraints on the field magnitude.}
The qubit Hamiltonian for EDSR is then given by 
\begin{multline}
H_q = H_0 ({\bf k} \rightarrow -i{\bm \nabla} - e\mathbf{A}/\hbar) + H_R ({\bf k} \rightarrow -i{\bm \nabla} - e\mathbf{A}/\hbar) \\
 + V_d(x, y) + (g_z/2) \, \mu_B B_z \sigma_z  + eE_\text{ac}x\cos (\omega t) \, \sigma_x , \label{eq:Hq}
\end{multline} 
where $\mathbf{A} = (B_z/2)(-y,x,0)$ and $g_z\approx 8$ is the Land\'e $g$ factor for Ge, in the direction perpendicular to the quantum well~\cite{Sammak2019}.
For a circular, parabolic dot, we assume an electrostatically defined confinement potential of the form 
\begin{equation}
V_d(x, y) = \frac{1}{2} m_x \omega_0^2 (x^2 + y^2),
\end{equation}
where $\hbar\omega_0$ is the energy splitting between the orbital levels when $B_z=0$.
If we now assume that $B_z>0$, but set $E_\text{ac}=0$, the eigenstates of $H_q$ are defined as Fock-Darwin orbitals~\cite{Fock1928,darwin1931}, for which the ground state ($n=0$) is given by 
\begin{equation}
    \phi_0(x,y) = \frac{1}{ a_0\sqrt{\pi}}\exp\left[-(x^2+y^2)/2a_0^2\right]~,
\end{equation}
and the first excited states ($n=1$) are given by 
\begin{equation}
\phi_{\pm 1}(x,y) = \frac{2}{ a_0^2\sqrt{\pi}}(x \pm iy)\exp\left[-(x^2+y^2)/2a_0^2\right] .
\end{equation}
For an out-of-plane magnetic field, we note that the dot is confined both electrostatically and magnetically, with an effective radius of $a_0=\sqrt{\hbar/|e B_z|}/(1/4+\omega_0^2/\omega_c^2)^{1/4}$, where $\omega_c=|e B_z|/m_x$ is the cyclotron frequency.

For hole-spin qubits, the logical (spin) states are formed exclusively within the ground-state orbital, $\phi_0$.
However, the EDSR spin-flip mechanism involves virtual transitions to $\phi_{\pm 1}$ via a second-order process that combines ac driving and SOC.
The driving term in Eq.~(\ref{eq:Hq}), $eE_\text{ac}x\cos (\omega t)$, is applied through one of the nearby top gates~\cite{kawakami14}, generating an orbital transition with $\Delta n=\pm 1$.
Initial proposals for hole-based EDSR~\cite{Bulaev:2007} therefore required Dresselhaus SOC, which can generate such $\Delta n=\pm 1$ transitions.
For group-IV materials, however, the Dresselhaus mechanism is normally absent, as pointed out above.
Moreover, the dominant $\alpha_{R2}$ term of the Rashba coupling, Eq.~(\ref{eq:HSO}), is cubic in $k$, as consistent with $\Delta n=\pm 3$, and therefore does not support EDSR.
An important conclusion of the present work is that the $\alpha_{R3}$ term, which is not typically considered in such calculations, provides the required $\Delta n=\pm 1$ transitions that support EDSR.
In what follows, we focus exclusively on this term.

\begin{figure}[t]
\includegraphics[width=\columnwidth]{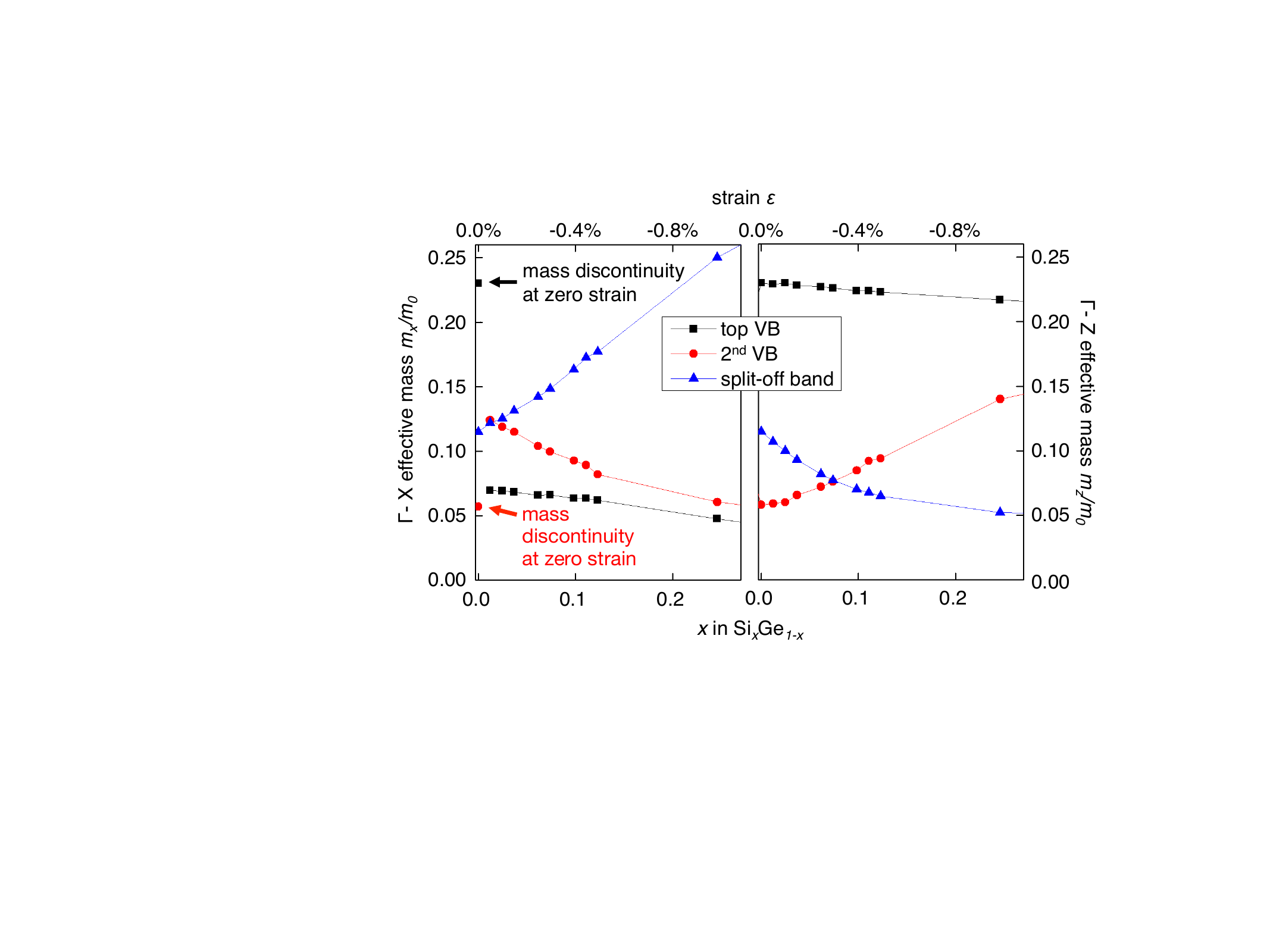}
\caption{Effective masses for the top three valence bands, in units of the free-electron rest mass $m_0$, obtained using DFT. 
\red{Here we consider a thin Ge well grown epitaxially on a relaxed SiGe alloy.
The resulting strain is uniaxial and compressive, and can be as large as 1\% along the growth axis. 
For a $[001]$ growth axis, the effective masses $m_z^*$ and $m_x^*$ are inequivalent.
While $m_z^*$ is found to vary smoothly with substrate composition, $m_x^*$ changes abruptly near $x\approx 0$, indicating an inversion of the band character: the top band becomes lighter than the second band, as consistent with Fig.~\ref{fig:old_fig1_bcde}(d), 
due to band hybridization. 
Such behavior can be explained by $\bf k\!\cdot\! p$ theory~\cite{bir1974}. 
Since the top two bands are no longer strictly light or heavy, we refer to them here as ``top" (or first) band and ``second" band.}
}
\label{dqdu3}
\end{figure}

To calculate the EDSR Rabi frequency $f_R$, we evaluate the full Hamiltonian, Eq.~(\ref{eq:Hq}), using the Fock-Darwin basis states, and perform a Schrieffer-Wolff transformation to eliminate the coupling to the excited states.
For resonant driving, with $\omega=\sqrt{\omega_0^2+\omega_c^2/4}$, we obtain
\begin{equation}
  \begin{array}[b]{rcl}
	h f_R &=& -\frac{e E_\text{ac} \alpha_{R3}}{2 a_0^2}\left[\left(\frac{1}{\Delta_1}+\frac{1}{\Delta_2} \right)-\left(\frac{1}{\Delta_3}+\frac{1}{\Delta_4}\right)\right] \\
	&~&- \frac{e^2 E_\text{ac} \alpha_{R3} B_z}{4 \hbar}\left[\left(\frac{1}{\Delta_1}+\frac{1}{\Delta_2} \right)+\left(\frac{1}{\Delta_3}+\frac{1}{\Delta_4}\right)\right],
	\label{eq:H_EDSR}
 \end{array}
\end{equation}
where 
\begin{equation}
\begin{array}[b]{rcl}
	\Delta_1 &\equiv& -\hbar \omega - \frac{1}{2} \hbar \omega_c  , \\
	\Delta_2 &\equiv& -\hbar \omega - \frac{1}{2} \hbar \omega_c - g_z \mu_B B_z  , \\
	\Delta_3 &\equiv& -\hbar \omega + \frac{1}{2} \hbar \omega_c + g_z \mu_B B_z ,\\
	\Delta_4 &\equiv&-\hbar \omega + \frac{1}{2} \hbar \omega_c .
\end{array}
\end{equation}
This result is explicitly proportional to $E_\text{ac}\alpha_{R3}$.
\red{Moreover, $f_R$ is found to be linear in $B_z$,
as readily verified by expanding Eq.~(\ref{eq:H_EDSR}) in powers of (small) $B_z$:
\begin{equation}\label{Rabi}
    |f_R| =  \frac{e E_{\text{ac}} \alpha_{R3} g_z \mu_B B_z m_x^2 a_0^2}{2\pi\hbar^5}.
\end{equation}
We note in Eq.~(\ref{Rabi}) that the Rabi frequency scales as $a_0^2$.
The explanation for this interesting behavior is that the EDSR strength is determined by the Rashba coupling between the ground and excited states of the  dot. 
Since larger quantum dots have smaller confinement energies, the excitation energies are also small, yielding faster EDSR.
In Sec.~\ref{sec:Rabi_results}, below, we provide numerical estimates for $f_R$, based on results of our DFT and $\bf k\!\cdot\! p$ calculations.}

\section{RESULTS}
\label{sec:results}

\red{We now describe the numerical results of our DFT and $\bf k\!\cdot\! p$ calculations.
We also discuss the shifts in energy caused by confinement and provide numerical estimates for Rabi frequencies that can be obtained from EDSR.
The main products of the DFT and $\bf k\!\cdot\! p$ calculations are the Ge band structures, as a function of strain.
The calculations also allow us to characterize the different bands, with regards to effective mass, spin, and band hybridization.
The results are summarized as follows.
In the limit of zero strain ($\varepsilon=0$), the topmost band is considered to be ``heavy," with a large transverse effective mass, $m_x$.
Away from $k=0$, this heavy band is doubly degenerate, with total spin quantum numbers $m_j=\pm 3/2$, which are well defined.
For nonzero strain, the top two bands become increasingly hybridized, with nonparabolic band structures.
Focusing mainly on the topmost band where the qubit is formed, $m_x$ abruptly jumps from being heavy to light.
This is the mass experienced by large dots (with small $k$), due to weak lateral confinement.
However small dots (with large $k$) may experience an effective mass that is heavy due to the band nonparabolicity.
Similarly, for large dots, the $m_j$ quantum number may have values near $\pm 3/2$.
However, for all $k^2=k_x^2+k_y^2>0$, the top two bands hybridize significantly, causing the $m_j$ quantum numbers to mix, such that $m_j$ is no longer a good quantum number.
Similar considerations also apply to the second valence band, although it does not house qubits.
For the reasons described above, we therefore adopt the labels ``top" (or ``first"), ``second," and ``split-off" for the three valence bands. 
Since they do not house qubits, the second and split-off bands are considered to be ``leakage" bands.}

\begin{figure}[t]
\includegraphics[width=2.6in]{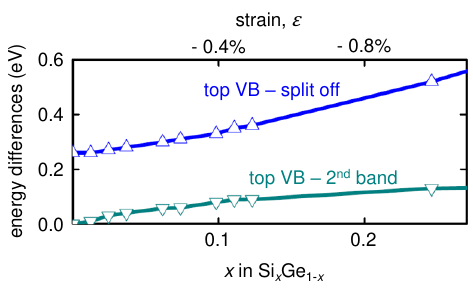}
\caption{Energy differences between the hole bands at the $\Gamma$ point as a function of the Si concentration in the substrate, $x$, obtained using DFT. Upward-pointing blue triangles correspond to the splitting between the top of the valence band and the split-off band, while downward-pointing teal triangles show the splitting between the first and second bands.}
\label{fig:dqdu2_dft}
\end{figure}

\begin{figure*}
\includegraphics[width=7in]{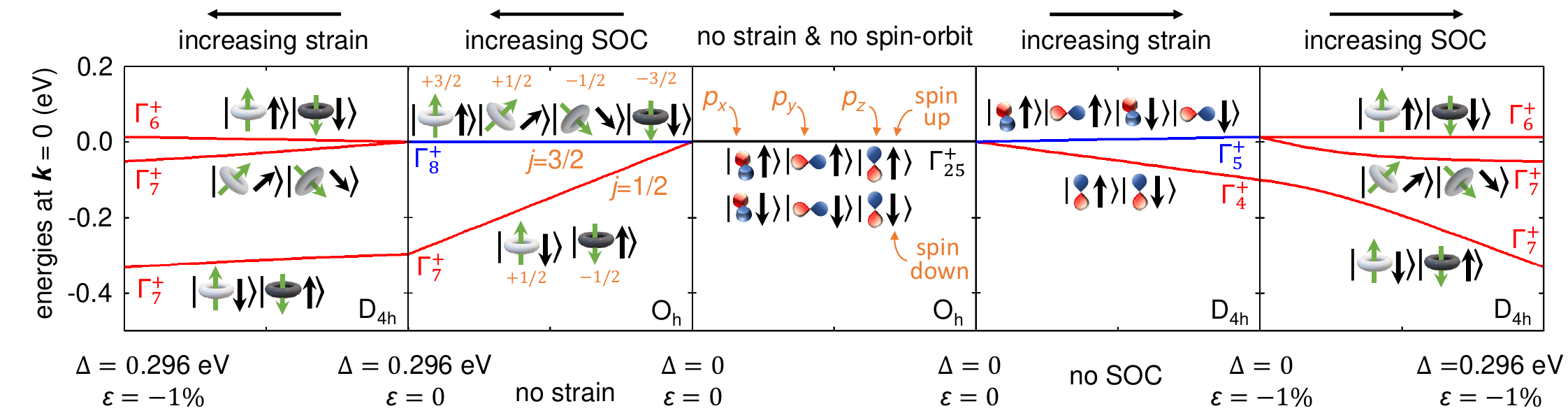}
\caption{
Energy levels calculated at the $\Gamma$ point, using the $\mathbf{k\!\cdot\!p}$ method, at zero magnetic field, which allows us to artificially decouple the effects of SOC (represented by the split-off band gap $\Delta$ of bulk Ge) and strain ($\varepsilon$). 
The five panels show results when these two parameters are independently varied between zero and their final values, corresponding to a strained quantum well with $x$=$0.25$. 
Level degeneracies are indicated by color: black for sixfold, blue for fourfold, and red for twofold.
The point symmetry groups and corresponding irreducible representations for the hole states are indicated in each case. 
The center panel represents the case with no SOC and no strain, in which the $p_x$, $p_y$, and $p_z$ orbitals and both spin states are  degenerate. Moving to the right, the strain is increased without including SOC, yielding a fourfold degenerate band spanned by $p_x$ and $p_y$, and a twofold degenerate $p_z$ band. Including SOC, the $p$ orbitals hybridize, creating states with different combinations of orbitals (represented now as tori) and spins, resulting in three doublets. 
Moving from the center panel to the left, including SOC but no strain yields a split-off, doubly-degenerate $j$=$1/2$ band and a fourfold degenerate $j$=$3/2$ band, as consistent with bulk, relaxed Ge at the $\Gamma$ point. We represent these states in a classical picture as having orbital angular momenta and spins either parallel (upper quadruplet) or anti-parallel (lower doublet). 
The different values of orbital angular momenta are represented by the colors of the orbital (darker tones for lower $m_j$) and by the inclination of the green vectors in relation to the vertical direction.
Including strain, the bands hybridize slightly such that $j$ and $m_j$ are no longer a good quantum number.
Here, the fully strained spectrum is identical to the far right-hand side of the figure.}
\label{fig:dqdu2}
\end{figure*}

\subsection{DFT Estimates}
\label{subsec:results_DFT}

DFT results are plotted in the main panels of Figs.~\ref{fig:old_fig1_bcde}(a) and \ref{fig:old_fig1_bcde}(b), where we compare Ge band structures for the cases of $x=0$ (unstrained Ge) and $x$=0.25 (strained Ge).
In the first case, the cubic symmetry ensures that the energy dispersion is identical for wavevectors in the plane of the quantum well ($k_x, k_y$) and the growth direction ($k_z$). 
(Here, the subscript $x$ refers to the [100] axis, rather than the alloy composition.)
In the second case, the X and Z points are inequivalent, as apparent in the figure. 
Focusing on holes, Figs.~\ref{fig:old_fig1_bcde}(c) and \ref{fig:old_fig1_bcde}(d) show blown-up views of the top of the valence band.
Since the quantum dot wave functions are constructed mainly from Bloch states at the very top of the band, the essential physics is captured in the band curvature at the $\Gamma$ point, which is proportional to the inverse effective mass.
In the case of strain, we observe anisotropic behavior in the $x$ (in-plane) and $z$ (out-of-plane) directions.
Figures~\ref{fig:old_fig1_bcde}(c) and \ref{fig:old_fig1_bcde}(d) also highlight the large energy splittings between the different bands under strain, which is key for defining the qubit states.

Figure~\ref{dqdu3} provides a more detailed picture of the in-plane ($m_x^*$) and out-of-plane ($m_z^*$) effective masses, obtained for strains in the range $\varepsilon\in [-1,0]$ percent.
The corresponding values of $x$ in the Si$_x$Ge$_{1-x}$ barrier alloy are also shown.
We note that the in-plane mass of the top two bands changes abruptly near $x$=$0$.
Remarkably, $m_x^*$ becomes lightest for the top band, over the experimental regime of interest ($x\gtrsim 0$), despite the usual label of ``heavy-hole" band.
As noted above, we therefore refrain from referring to heavy or light holes in this work, adopting instead the terminology ``first" (or ``top"), and ``second" bands.
For $m_z^*$, the top band remains heaviest for all $x$ considered here, and is a smooth function of the strain.
These results are in reasonable agreement with several recent experiments~\cite{dobbie12,scappucci2018,Morrison:2016}, and they agree very well with Ref.~\cite{Lodari2019}, in which band nonparabolicity is explicitly accounted for.

Figure~\ref{fig:dqdu2_dft} shows the corresponding results for the energy dispersion of the valence-band edges.
In the limit $x\rightarrow 0$, the top two bands become degenerate, and the split-off band is lower in energy by an amount $\Delta=0.29$~eV, which compares well with the experimentally measured value of 0.296~eV~\cite{schaffler97}.
For $x>0$, the band degeneracy is lifted by a significant amount, of order 100~meV for typical quantum-well heterostructures.
In contrast with the effective mass, no abrupt change occurs for the valence-band edges near $x$=0.

To summarize the present results, DFT predicts a sudden change in the in-plane mass of the top band as the strain decreases from zero, with $m_x$ becoming very light.
Moreover, the degeneracy of the top two bands is lifted, and the energy splitting between all the bands is enhanced.
These results are all consistent with recent experiments.

\subsection{$\mathbf{k\!\cdot\!p}$ Analysis}
\label{subsec:results_k_dot_p}

The $\mathbf{k\!\cdot\!p}$ approach allows us to explore the mechanisms that cause the changes in the band structure and clarify their separate roles.
In Fig.~\ref{fig:dqdu2}, we plot the edges of the top three valence bands, as a function of either strain or SOC. \red{The symmetries of each band are indicated for the $\Gamma$ point.}
By following the progression from a single sixfold-degenerate band (center panel) to three twofold-degenerate bands (outer panels), we infer that the splitting of the top two bands requires both strain and SOC. 
\red{The resulting top valence band is two-fold degenerate, in accordance with time-reversal symmetry, and can be split by an external magnetic field to define the qubit states, $|0\rangle$ and $|1\rangle$, and their quantization axis.}
The calculations also show that the hybridization of the topmost bands occurs at second order, via strain-induced coupling to the split-off band. 
Since this effect is weak, the total angular momentum in the top band, which defines the qubit, is still given by $j\approx 3/2$ and $m_j\approx \pm 3/2$ to a reasonable approximation, as indicated in the figure. 
Spin flips with $\Delta m_s=\pm 1$ are allowed by EDSR, however, via the Rashba coupling mechanism described above.

In the $\mathbf{k\!\cdot\!p}$ calculations, we note that strain has been introduced perturbatively.
Hence, although the energy splitting of the lowest valence band is accurate when $\varepsilon=0$, since it is taken as an input parameter, the calculated energies become increasingly inaccurate for higher strain values. 
For example, when $\varepsilon=-1$\%, the more accurate DFT result of $\Delta$=0.53~eV is $>$50\% larger than the $\mathbf{k\!\cdot\!p}$ estimate.
Likewise, the  $\mathbf{k\!\cdot\!p}$ energy splitting of 0.06~eV between the top two valence bands is approximately half the DFT estimate of 0.13~eV.

To summarize, the $\mathbf{k\!\cdot\!p}$ theory reproduces the general features of the band structure that was obtained more rigorously using DFT. 
Although $\mathbf{k\!\cdot\!p}$ methods are less accurate than DFT, they allow us to clarify that both strain and SOC are required to fully lift the band degeneracy at $\mathbf{k}=0$.

\subsection{Quantum Well Corrections to the Energy}

The energies plotted in Fig.~\ref{fig:dqdu2_dft} were obtained without including the quantum-well subband confinement energies, which differ for different bands, and can be sizeable.
Here we show that the subband contribution to the hole energy does not compromise the energy splitting between the top two valence bands or change the effective ordering between them.

\begin{figure}
	\includegraphics[width=2.4in]{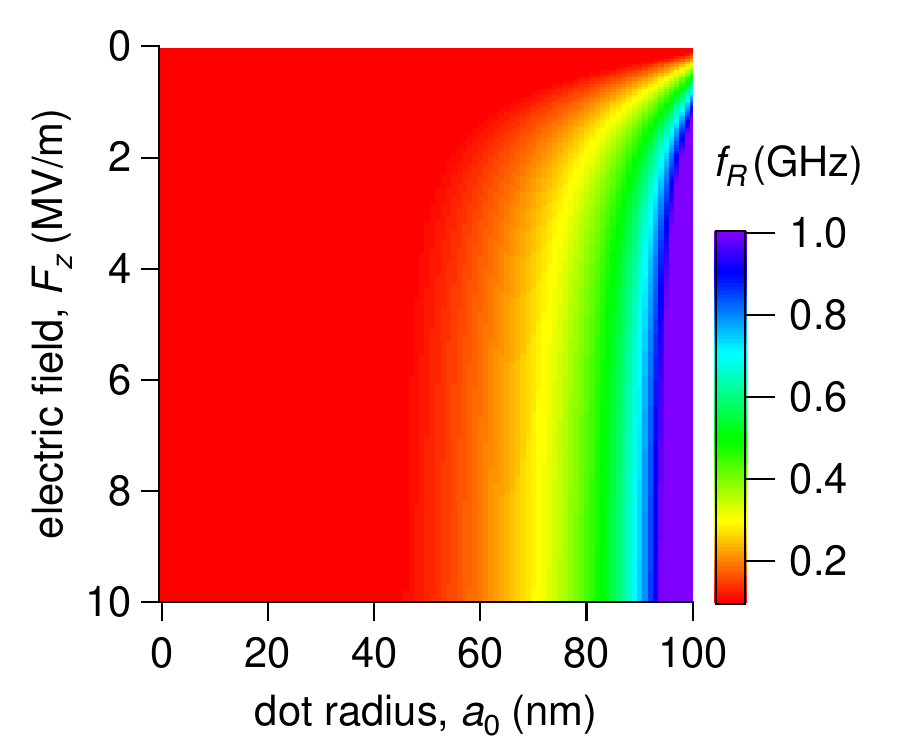}
	\caption{Color map of the EDSR Rabi frequency, $f_R$, as a function of both the vertical electric field, $F_z$, and the effective dot radius, $a_0$, with magnetic field $B_z$=0.06~T, quantum well width $d$=20~nm, and microwave driving amplitude~\cite{Salfi2016} $E_\text{ac}$=0.1~MV/m. All materials parameters assume a Si concentration of $x$=0.25 in the barrier alloy.}
	\label{fig:color_maps}
\end{figure}

The subband energies differ for the top two valence bands due to their different effective masses.
We can estimate these effects by assuming a triangular, vertical confinement potential, as in Eq.~(\ref{eq:triangle}).
Here, we assume an electric field value of $F_z\approx ep/\epsilon$, which is the field required to accumulate a 2D hole gas with density $p=4\times 10^{11}$~cm$^{-2}$, and we linearly interpolate the dielectric constant in the Si$_x$Ge$_{1-x}$ barrier layer, obtaining the relation $\epsilon(x)$=$(16.2-4.5x)\epsilon_0$, where $\epsilon_0$ is the vacuum permittivity.
We further assume that the vertical extent of the wave function is less than the quantum well width, allowing us to ignore the bottom edge of the well. 
The triangular potential has known solutions~\cite{daviesbook}, yielding a confinement energy of $2.34~E_0$ for the first subband and $4.09~E_0$ for the second subband, where $E_0$=$(\hbar^2e^4p^2/2m_z^*\epsilon^2)^{1/3}$ is a characteristic energy scale and $m_z^*$ depends on both the alloy composition and the particular valence band. 
(Note that we do not consider band-nonparabolicity effects here, although they can be significant due to the large energies involved.)
In this way, when $x$=$0.25$, we obtain a total energy splitting (including both band and subband energies) of 140~meV for the lowest-energy confined holes in the first and second valence bands, with the first band still having the lowest energy. 
In comparison, the energy splitting between the first and second subbands within the top valence band, is 27.7~meV, which therefore represents the predominant leakage channel for the qubit. 
We conclude that band and subband excitations of the qubit level are much larger than other relevant energy scales in this system, including the thermal energy of the hole reservoirs (5-15~$\mu$eV), the inter-dot tunnel couplings ($\sim$200~$\mu$eV), and exchange interactions ($\sim$200~$\mu$eV). 

\subsection{Rabi Frequency Estimates}
\label{sec:Rabi_results}
\begin{figure}[t]
\vspace{0.in}
		\includegraphics[width=2.5in]{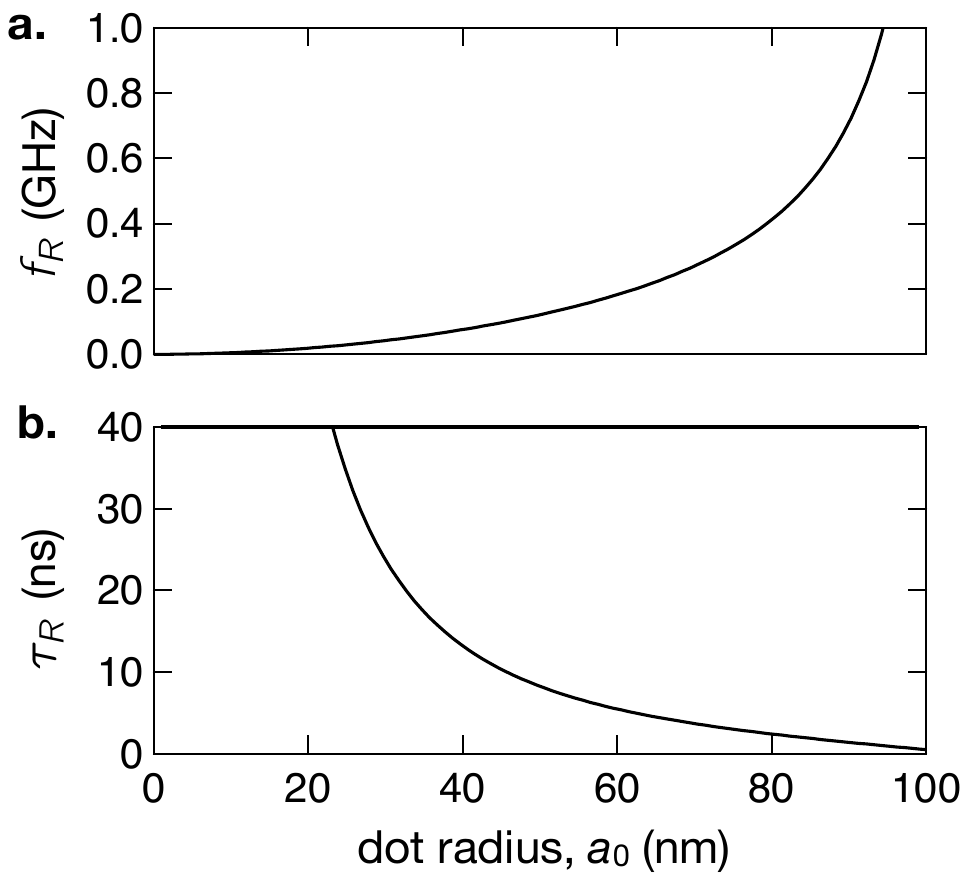}
		\caption{Calculated values of (a) the EDSR Rabi frequency, $f_R$, and (b) the corresponding $X_{2\pi}$ gate time, $\tau_R$=$1/f_R$, as a function of the effective dot radius $a_0$.
		Here, the simulation parameters are the same as in Fig.~\ref{fig:color_maps}, with $F_z = 4.5$ MV/m.} 
		\label{fig:line_cut}
\end{figure}

In Sec.~\ref{sec:Rabi_calculations}, particularly in Eq.~(\ref{eq:H_EDSR}), we obtained general results for the EDSR Rabi frequency $f_R$ as a function of system parameters.
In Fig.~\ref{fig:color_maps}, we now plot the dependence of the Rabi frequency on the dot radius $a_0$ and the vertical electric field $F_z$. 
In Fig.~\ref{fig:line_cut}, we further show a line-cut through this data, and a corresponding plot of $\tau_R=1/f_R$, representing the gate time for an $X_{2\pi}$ gate operation.
Generally, we find that larger dots yield faster gate operations due to their smaller orbital energies.
(We note that, for sufficiently large $a_0$, the perturbative methods used here become inaccurate.)
\red{To take an example, for a vertical field of $F_z=4.8$~MV/m, which is typical for some experiments but can be as large as 10~MV/m~\cite{delVecchio:2020}}, and effective dot radii in the range of 30-60~nm \cite{Sammak2019,scappucci2018,Hendrickx2020}, Rabi frequencies can be of order 0.2~GHz, corresponding to a 5~ns gate time for an $X_\pi$ gate.
Such fast gates are very promising for high-fidelity quantum gate operations.

\section{\textbf{DISCUSSION and Conclusions}}
\label{sec:discussion}
Recent experimental work has already demonstrated that holes in germanium are promising as qubits.  
In this work, we have explored how confinement and strain are critical for achieving such strong performance, particularly in the context of EDSR-based gate operations.
We have also demonstrated that operating the qubits in an out-of-plane magnetic field may be advantageous because of the highly anisotropic $g$-factor.

To conclude, we comment on the expected decoherence mechanisms affecting Ge hole spins.  
As mentioned in the introduction, hyperfine interactions are suppressed for hole spins due to the $p$-orbital character of the valence band~\cite{Bulaev:2007,de11}, and the low natural abundance ($<$8\%) of Ge isotopes with nonzero nuclear spin, which can be further reduced by isotopic purification~\cite{hu07}.
However, charge noise is ubiquitous in semiconductor devices~\cite{dial13}, including Ge quantum dots, particularly in the vicinity of the gate oxides.  
Although the poor quality of Ge oxides could exacerbate this problem, the simple inclusion of a Si capping layer should bring Ge/SiGe on par with related systems, such as Si-based qubits.
Similarly, phonon noise should be similar in Ge and Si-based devices; in both cases, phonon effects are much weaker than in GaAs charge~\cite{hayashi03} or spin qubits~\cite{hu11,gamble12}, due to the absence of piezoelectric phonons.
Hence, hole spins in Ge quantum wells should be relatively well protected from their environment, making them particularly strong candidates for quantum dot qubits.

\section*{ACKNOWLEDGEMENTS}

We are grateful to D.~J.~Paul, G.~Scappucci, D.~E.\ Savage, O. P. Sushkov, D. S. Miserev and M. A. Eriksson for illuminating discussions and helpful information. 
Work in Brazil was performed as part of the INCT-IQ and supported by Centro Nacional de Processamento de Alto Desempenho em São Paulo (CENAPAD-SP, project UNICAMP/FINEP-MCT), CNPq (304869/2014-7, 308251/2017-2, and 309861/2015-2), and FAPERJ (E-26/202.915/2015 and 202.991/2017). 
Work in Australia was supported by the Australian Research Council Centre of Excellence in Future Low-Energy Electronics Technologies (project CE170100039). 
Work in the U.S.A. was supported in part by ARO (W911NF-17-1-0274 and W911NF-17-1-0257), NSF (OISE-1132804), and the Vannevar Bush Faculty Fellowship program sponsored by the Basic Research Office of the Assistant Secretary of Defense for Research and Engineering and funded by the Office of Naval Research through grant N00014-15-1-0029.  
M.\ F.\ also acknowledges support from the NSF Quantum Leap Challenge Institute for Hybrid Quantum Architectures and Networks (NSF Award 2016136).
The views and conclusions contained in this document are those of the authors and should not be interpreted as representing the official policies, either expressed or implied, of the Army Research Office (ARO), or the U.S. Government. The U.S. Government is authorized to reproduce and distribute reprints for Government purposes notwithstanding any copyright notation herein.

% \vspace{.2in}\noindent{\textbf{COMPETING INTERESTS}}

% \noindent
% The authors declare no conflict of interest.\vspace{.2in}

% \noindent{\textbf{CONTRIBUTIONS}}

% \noindent
% LAT performed the DFT calculations under the supervision of RBC. EM performed the EDSR calculations under the supervision of ARH and DC. SNC, MF, XH, BK, and ALS conceived the project. All authors discussed the results and contributed to the preparation of the manuscript.

\bibliography{saraiva}

\end{document}